\documentstyle[12pt]{article}
\input epsf

\def\beq{\begin{equation}}
\def\eeq{\end{equation}}

\def\al{\alpha}
\def\bt{\beta}
\def\ga{\gamma}

\def\de{\delta}
\def\De{\Delta}

\def\Si{\Sigma}

\def\lam{\lambda}

\def\ep{\epsilon}

\def\l{\left (}
\def\r{\right )}
\def\lx{\left |}
\def\rx{\right |}

\def\fr{\frac}
\def\la{\label}
\def\hs{\hspace}

\def\ran{\rangle}
\def\lan{\langle}

\begin{document}

\begin{titlepage}
\begin{flushright}
BA-01-21\\
\end{flushright}

\begin{center}
{\Large\bf   Realistic Supersymmetric $SU(6)$
}
\end{center}
\vspace{0.5cm}
\begin{center}
{\large Qaisar Shafi$^{a}$\footnote {E-mail address:
shafi@bartol.udel.edu} {}~and
{}~Zurab Tavartkiladze$^{b}$\footnote {E-mail address:
z\_tavart@osgf.ge} }
\vspace{0.5cm}

$^a${\em Bartol Research Institute, University of Delaware,
Newark, DE 19716, USA \\

$^b$ Institute of Physics, Georgian Academy of Sciences,
380077 Tbilisi, Georgia}\\

\end{center}

\vspace{1.0cm}

\begin{abstract}
We present an example of SUSY $SU(6)$ GUT, which predicts an excellent
value for $\al_s(M_Z)(\simeq 0.119)$, in comparison with the
value $\al_s^0(M_Z)\simeq 0.126$ of minimal SUSY $SU(5)$.
A crucial role is played by the vectorlike multiplets from the matter
sector, 
whose masses
lie below the GUT scale. For a
realistic pattern of fermion masses, the adjoint scalar of $SU(6)$ has VEV
along the 
$SU(4)\times SU(2)\times U(1)$ direction. This also offers a natural resolution
of the doublet-triplet (DT) splitting through the
pseudo Goldstone Boson mechanism.

\end{abstract}

\end{titlepage}

The minimal SUSY $SU(5)$ model suffers from a variety of nagging problems.
For instance, the measured value of the strong coupling 
$\al_s(M_Z)=0.119\pm 0.002$ \cite{data},
while the predicted value is
$\al_s^0=0.126$ \cite{alpsu5}. 
It also predicts the wrong asymptotic relations
$m_s^{(0)}=m_{\mu }^{(0)}$, 
$m_d^{(0)}/m_s^{(0)}=m_e^{(0)}/m_{\mu }^{(0)}$.
And finally, although SUSY guarantees stability of scales against
radiative
corrections, the origin of DT splitting remains unexplained. 
In attempting to resolve these problems, one can either consider some
extended versions of $SU(5)$ or an alternative GUT scenario. In fact, for
obtaining
a desirable value of $\al_s(M_Z)$, some additional states below the GUT
scale could play an important role \cite{impal1, impal2}.
For realistic fermion masses, either a scalar $45$ plet
\cite{45pl} or additional fermionic states \cite{impal2} can be
introduced.
Within $SU(5)$, solution of the DT splitting problem requires a rather
complicated $(50+\overline{50}+75)$ set of scalars, which turn out to be 
crucial for
realization of the missing partner mechanism \cite{mispart}. Replacing $SU(5)$
with
$SO(10)$, one can achieve DT splitting through the missing VEV
mechanism \cite{VEVso}\footnote{See \cite{VEVsu} for
examples of 
missing VEV solutions in $SU(N)$.}.
Very attractive and promising scenarios are those in which the light higgs
doublets emerge as pseudo-Goldstone Bosons (PGB). This idea is
easily realized within $SU(6)$ \cite{pgbsu61, pgbsu62}, 
$SU(3)\times SU(3)\times SU(3)$ \cite{ananth, pgbsu3} or
flipped $SU(6)$ \cite{flipsu6} models. Also, $SU(6)$ scenarios with
additional
custodial symmetries can provide a natural understanding of DT splitting 
\cite{cus}.

In this letter we show how these three problems could be 
simultaneously resolved by considering an $SU(6)$ GUT. The
value of $\al_s(M_Z)$, it turns out, is closely tied with the matter sector,
and is expressed
through some asymptotic mass relations.
It is interesting to note that a realistic
pattern of fermion masses unequivocally requires the VEV of the adjoint higgs
to
be along the $SU(4)\times SU(2)\times U(1)$ direction. This also permits
realization of the PGB mechanism \cite{pgbsu61, pgbsu62, ananth}
for achieving a natural DT splitting.

\vspace{0.5cm}

Consider the SUSY $SU(6)$ GUT with chiral `matter' multiplets 
$15+\bar 6+\bar 6 \hs{0.4mm}'$ per generation.
In terms of $SU(5)$: $15=10+5$, $\bar 6=\bar 5+1$ (and same for 
$\bar 6 \hs{0.4mm}'$). Thus, we have the additional $SU(5)$  
$\bar 5+5$ vectorlike states, which decouple after $SU(6)$ breaking.
At first glance, since they are complete $SU(5)$ plets, one may think that
the picture of gauge coupling unification will not be altered at one
loop level. However, it turns out that the doublet and triplet fragments from
these additional $5$ ($\bar 5$) plets are split in mass. 
This happens because, in order to get a realistic pattern of down quark and
charged lepton masses, we somehow must remove the degeneracy
between their mass matrices. If this is done, then the heavy
vectorlike doublet and triplet states also will acquire different masses,
and their ratios will be
expressed through asymptotic mass relations of down quarks and charged
leptons, giving rise to the possibility of predicting $\al_s(M_Z)$.

The relevant $SU(6)$ invariant couplings, in lowest order, are of the form
$15(\bar 6+\bar 6\hs{0.4mm}')\bar H$, where
$\bar H$ ($H$) is an antisextet (sextet) scalar field. In order to avoid
the wrong
asymptotic relations $m_s^{(0)}=m_{\mu }^{(0)}$, 
$m_d^{(0)}/m_s^{(0)}=m_e^{(0)}/m_{\mu }^{(0)}$ we will insert in these
couplings 
the $SU(6)$ adjoint scalar $\Si (35)$ [this can be realized through a 
$Z_2$ symmetry $\Si \to -\Si $, 
$(\bar 6, \bar 6\hs{0.4mm}')\to -(\bar 6, \bar 6\hs{0.4mm}')$]. For a
transparent demonstration, let us first consider the case of one
generation. The relevant couplings are:

\beq
\fr{1}{M} 15_{ij}\Si^{\hs{0.4mm}i}_m\l \al \hs{0.4mm}
\bar{6}^{\hs{0.4mm}m}\bar
H^{\hs{0.4mm}j}+\bt \hs{0.4mm}\bar 6^{\hs{0.4mm}j}\bar H^{\hs{0.4mm}m}+
\al \hs{0.4mm}' \hs{0.4mm}\bar 6'^{\hs{0.4mm}m}\bar
H^{\hs{0.4mm}\hs{0.4mm}j}+
\bt \hs{0.4mm}'\hs{0.4mm} \bar 6'^{\hs{0.4mm}j}\bar H^{\hs{0.4mm}m}\r~,
\la{yuk}
\eeq
where $i, j, m$ are $SU(6)$ indices, $\al, \dots , \bt \hs{0.4mm}'$ -
are dimensionless couplings, and $M$ is some cutoff mass scale. $\Si $
and $\bar H$ have VEVs of the same order ($\sim M_G$), and the light higgs
doublet $h_d$ is suppressed by equal weights in these plets. It is
easy to verify that the relevant terms are built with the higgs
doublet $h_d$ extracted from $\bar H$, and we will ignore terms in which
the doublets from $\Si $ participate (such terms do not lead to
light fermion masses to be identified as
quarks and leptons, will couple with decoupled states).
{}From (\ref{yuk}), we have:

\begin{equation}
\begin{array}{cc}
 & {\begin{array}{cc}
\!\!\!d^{\hs{0.2mm}c}~~~~~~~~~~~~~~~~
&\,\,~~~d^{\hs{0.2mm}c}\hs{0.4mm}'~~~~
\end{array}}\\ \vspace{2mm}\hat{M}_D=
\begin{array}{c}
q
\\
\\
\bar d^{\hs{0.2mm}c} 

\end{array}\!\!\!\!\!\!\!\! &{\left(\begin{array}{cc}
\!\!\!\, (\al \Si_c-\bt \Si_w)h_d~~,  
&\!\!\!\!\, (\al\hs{0.4mm}'\hs{0.4mm} \Si_c-
\bt \hs{0.4mm}'\hs{0.4mm} \Si_w)h_d
\\
\\
\!\!\!\, (\al \Si_c-\bt \Si_6)v~~,  
&\!\!\!\!\, (\al\hs{0.4mm}'\hs{0.4mm} \Si_c-
\bt \hs{0.4mm}'\hs{0.4mm} \Si_6)v

\end{array}\!\!\right)\!\fr{1}{M} }~,
\end{array}  \!\!~~~
\la{matd}
\eeq

\begin{equation}
\begin{array}{cc}
 & {\begin{array}{cc}
\!\!\!l~~~~~~~~~~~~~~~~
&\,\,~~~l\hs{0.4mm}'~~~~
\end{array}}\\ \vspace{2mm}\hat{M}_E=
\begin{array}{c}
e^c
\\
\\
\bar l

\end{array}\!\!\!\!\!\!\!\! &{\left(\begin{array}{cc}
\!\!\!\, (\al -\bt )\Si_wh_d~~,
&\!\!\!\!\, (\al\hs{0.4mm}'\hs{0.4mm} -
\bt \hs{0.4mm}'\hs{0.4mm} )\Si_wh_d
\\
\\
\!\!\!\, (\al \Si_w-\bt \Si_6)v~~,
&\!\!\!\!\, (\al\hs{0.4mm}'\hs{0.4mm} \Si_w-
\bt \hs{0.4mm}'\hs{0.4mm} \Si_6)v
  
\end{array}\!\!\right)\!\fr{1}{M} }~,
\end{array}  \!\!~~~
\la{matl}
\eeq
where $15\supset (q, e^c, \bar l, u^c)$, $\bar 6\supset (d^c, l)$,
$\bar 6\hs{0.4mm}'\supset (d^c\hs{0.4mm}', l\hs{0.4mm}')$, and for the scalar
VEVs $\lan \bar H\ran \equiv v $, 
$\lan \Si \ran ={\rm Diag}(\Si_c, \Si_c, \Si_c, \Si_w, \Si_w,
\Si_6)$, with $\Si_6=-3\Si_c-2\Si_w$. {}From (\ref{matd}),
(\ref{matl}) we see that pairs of doublet and triplet states decouple
with masses $\sim v\lan \Si \ran /M$, while the light down quark and charged
lepton's masses are $\sim h_d\lan \Si \ran /M$. More precisely, from
(\ref{matd}), (\ref{matl}),

$$
{\rm Det}(\hat{M}_D)=\fr{v}{M^2}\Si_c(\Si_w-\Si_6)
(\al \bt \hs{0.4mm}'-\bt \al \hs{0.4mm}' )h_d~,
$$
\beq
{\rm Det}(\hat{M}_E)=\fr{v}{M^2}\Si_w(\Si_w-\Si_6)
(\al \bt \hs{0.4mm}'-\bt \al \hs{0.4mm}' )h_d~.
\la{dets}
\eeq
{}From (\ref{dets}) [and also from (\ref{matd}), (\ref{matl})] it is
obvious
that the symmetry breaking patterns 
$SU(5)\times U(1)$ and $SU(3)\times SU(3)\times U(1)$ are not plausible,
since, in these cases, we either have degeneracy between $\hat{M}_D$
and $\hat{M}_E$ or the determinants in (\ref{dets}) are zero [in the latter
case
some quark and lepton states are massless]. We therefore conclude that the
only possible $\lan \Si \ran $ VEV which can lead to a realistic fermion mass
pattern is
$SU(4)\times SU(2)\times U(1)($e.g. $\Si_c=\Si_6=-\Si_w/2$)

\beq
\lan \Si \ran ={\rm Diag}(1, 1, 1, -2, -2, 1)\cdot V~.
\la{421dir}
\eeq
Using (\ref{421dir}) in (\ref{dets}), one obtains

\beq
{\rm Det}( \hat{M}_E) =-2\cdot{\rm Det}( \hat{M}_D) ~,
\la{reldet}
\eeq
which implies $m_e^{(0)}M^l=2m_d^{(0)}M^{d^c}$, where $M^l$, $M^{d^c}$
are the masses of the heavy doublet and triplet components respectively, and
$m_e^{(0)}$, $m_d^{(0)}$ denote the asymptotic values of charged lepton and
down
quark mases. Therefore,

\beq
\fr{M^l}{M^{d^c}}=2\fr{m_d^{(0)}}{m_e^{(0)}}~.
\la{relmass}
\eeq
Knowing the asymptotic value of $\fr{m_d^{(0)}}{m_e^{(0)}}$ for a
given generation, we calculate through (\ref{relmass}) the ratio
$M^l/M^{d^c}$. The latter give us possibility to predict the value of 
$\al_s(M_Z)$.

Analagous results can be obtained for the case with three generations, and
as we will see, even inclusion of intergeneration mixings do not modify the
picture.  
Instead of (\ref{matd}),
(\ref{matl}) we will have $6\times 6$ matrices. Using (\ref{421dir}),
the appropriate mass matrices are:

\begin{equation}
\begin{array}{cc}
 & {\begin{array}{cc}
~~&\,\,~~~~~
\end{array}}\\ \vspace{2mm}
\begin{array}{c}
\\  \\ 
\end{array}\hs{-1mm}\!\!\!\!\!\!\!\! &{\hat{\cal M}_D\!=\!\!
\left(\begin{array}{cc}
\!\!\!\, (\hat{\al} \!+\!2\hat{\bt} )h_d~\hs{0.3mm},
&\!\!\!\!\, (\hat{\al}\hs{0.4mm}'\hs{0.4mm}\!+\!
2\hat{\bt} \hs{0.4mm}'\hs{0.4mm} )h_d
\\
\\
\!\!\!\, (\hat{\al} \!-\! \hat{\bt} )v~~,
&\!\!\!\!\, (\hat{\al}\hs{0.4mm}'\hs{0.4mm}\!-\!
\hat{\bt} \hs{0.4mm}'\hs{0.4mm} )v
\end{array}\!\!\right)\!\fr{V}{M} }\hs{0.2mm},\!

\end{array}  \!\!
\begin{array}{cc}
 & {\begin{array}{cc}
~&\,\,
~~~~~~
\end{array}}\\ \vspace{2mm}
\begin{array}{c}
 \\ 
\end{array}\hs{-1mm}\!\!\!\!\!\!\!\!\!\! &{\hat{\cal M}_E\!=\!\!
\left(\begin{array}{cc}
\!\!\!\, 2(\hat{\bt} \!-\!\hat{\al} )h_d~\hs{0.5mm},
&\!\!\!\!\, 2(\hat{\bt}\hs{0.4mm}'\hs{0.4mm}\!-\!
\hat{\al} \hs{0.4mm}'\hs{0.4mm} )h_d
\\
\\
\!\!\!\, -\!(2\hat{\al} \!+\!\hat{\bt} )v~~,
&\!\!\!\!\, -\!(2\hat{\al}\hs{0.4mm}'\hs{0.4mm}\!+\!
\hat{\bt} \hs{0.4mm}'\hs{0.4mm} )v
\end{array}\!\!\right)\!\fr{V}{M} }\hs{0.4mm},


\end{array}~~~
\label{bigmats}
\eeq
where $\hat{\al }, \dots ,\hat{\bt}\hs{0.4mm}'$ indicate $3\times 3$
matrices in generation space.

It is not difficult to find a relation
between the determinants of matrices in (\ref{bigmats}). Recall that
determinants remain unchanged by making some linear manipulations with 
their rows and coulomns. More precisely:

\begin{equation}
\begin{array}{cc}
 & {\begin{array}{cc}
~~&\,\,~~~~~
\end{array}}\\ \vspace{2mm}
\begin{array}{c}
\\  \\
\end{array}\hs{-1mm}\!\!\!\!\!\!\!\! &{{\rm Det}( \hat{\cal M}_D) \!=\!
\lx \begin{array}{cc}
\!\!\!\, (\hat{\al} \!+\!2\hat{\bt} )h_d~\hs{0.3mm},
&\!\!\!\!\, (\hat{\al}\hs{0.4mm}'\hs{0.4mm}\!+\!
2\hat{\bt} \hs{0.4mm}'\hs{0.4mm} )h_d
\\
\\
\!\!\!\, (2\hat{\al} \!+\! \hat{\bt} )v~~,
&\!\!\!\!\, (2\hat{\al}\hs{0.4mm}'\hs{0.4mm}\!+\!
\hat{\bt} \hs{0.4mm}'\hs{0.4mm} )v
\end{array}\!\!\rx \!\l \fr{V}{M}\r ^6 }\!\!\!

\end{array}  \!\!
\begin{array}{cc}
 & {\begin{array}{cc}
~&\,\,
~~~~~~
\end{array}}\\ \vspace{2mm}
\begin{array}{c}
 \\
\end{array}\hs{-1mm}\!\!\!\!\!\!\!\!\!\! &{\!\!\!=\!
\lx \begin{array}{cc}
\!\!\!\, (\hat{\bt} \!-\!\hat{\al} )h_d~\hs{0.5mm},
&\!\!\!\!\, (\hat{\bt}\hs{0.4mm}'\hs{0.4mm}\!-\!
\hat{\al} \hs{0.4mm}'\hs{0.4mm} )h_d
\\
\\
\!\!\!\, (2\hat{\al} \!+\!\hat{\bt} )v~~,
&\!\!\!\!\, (2\hat{\al}\hs{0.4mm}'\hs{0.4mm}\!+\!
\hat{\bt} \hs{0.4mm}'\hs{0.4mm} )v
\end{array}\!\!\rx \!\l \fr{V}{M}\r^6 }\hs{0.4mm},


\end{array}~~~
\label{bigdet}
\eeq
Comparing the last determinant in (\ref{bigdet}) with the second matrix in
(\ref{bigmats}), we see that

\beq
{\rm Det}( \hat{\cal M}_E) =-8\cdot{\rm Det}( \hat{\cal M}_D) ~.
\la{relbigdets}
\eeq
Therefore,
$m_e^{(0)}m_{\mu }^{(0)}m_{\tau }^{(0)}M^l_1M^l_2M^l_3=
8m_d^{(0)}m_s^{(0)}m_b^{(0)}M^{d^c}_1M^{d^c}_2M^{d^c}_3$,
where $M^l_i$, $M^{d^c}_i$ denote the masses of heavy doublets and triplets of
the corresponding generation. Finally:

\beq
\fr{M^l_1M^l_2M^l_3}{M^{d^c}_1M^{d^c}_2M^{d^c}_3}=
8\l \fr{m_dm_sm_b}{m_em_{\mu }m_{\tau }}\r^{(0)}~.
\la{rel3mass}
\eeq
We will see below that the value of $\al^{-1}_s(M_Z)$ will depend
logarithmically on the ratio in
(\ref{rel3mass}).

The solutions of the RGEs are \cite{alpsu5}:

\beq
\al^{-1}_G=\al^{-1}_a-\fr{b_a}{2\pi }\ln \fr{M_G}{M_Z}-
\fr{b^l_a}{2\pi }\Si_i \ln \fr{M_G}{M^l_i}-
\fr{b^{d^c}_a}{2\pi }\Si_i \ln \fr{M_G}{M^{d^c}_i}+
\De_a+\de_a~,
\la{rge}
\eeq
where $\al_G$ is the gauge coupling at the GUT scale, $\al_a$ the
gauge coupling at $M_Z$ ($\al_{1, 2, 3}$ are gauge couplings of
$U(1)$, $SU(2)_W$ and $SU(3)_c$ respectively), while

\beq
b_a=(\fr{33}{5}, 1, -3)~,~~
b^l_a=(\fr{3}{5}, 1, 0)~,~~
b^{d^c}_a=(\fr{2}{5}, 0, 1)~.
\la{bfacs}
\eeq
The $\De_a$ include all possible threshold corrections and two loop
effects of MSSM. $\de_a$ denote the difference between MSSM and the present model
of the gauge coupling running
from $M_1^{d^c}$ (lowest possible 
intermediate scale) up to $M_G$ in two loop approximation,

\beq
\de_a=\fr{1}{4\pi }\l \fr{b_{ab}^{\rho }}{b_b^{\rho }}
\ln \fr{\al_b(M_{\rho })}{\al_b(M_{\rho+1})}-\fr{b_{ab}}{b_b}
\ln\fr{\al_b(M_1^{d^c})}{\al_G^0}
\r ~,
\la{twoloop}
\eeq
where summation over $\rho $ and $b$ indices is implied. $\rho $
enumerates the heavy vectorlike doublet and triplet states below the
GUT scale, and $M_{\rho }$ and $b^{\rho }_a$, $b^{\rho }_{ab}$ are the corresponding
mass scale and b-factors (which depend on energy scale ) respectively.
$b_{ab}$ denote two loop b-factors of MSSM. In (\ref{twoloop}) the
appropriate
couplings are calculated in one loop approximation. $\al_G^0$ is the gauge
coupling  of MSSM at $M_G$.

For the time being in (\ref{rge}) we will ignore $\de_a$.
Calculating the combination $12\al^{-1}_2-5\al^{-1}_1-7\al^{-1}_3$ and taking
into account (\ref{rel3mass}), one obtains:

\beq
\l \al^{-1}_s\r'=\l \al^{-1}_s\r^0+\fr{9}{14\pi}\ln
\fr{M^l_1M^l_2M^l_3}{M^{d^c}_1M^{d^c}_2M^{d^c}_3}=\l \al^{-1}_s\r^0+
\fr{9}{14\pi}\ln \l 8\fr{m_dm_sm_b}{m_em_{\mu }m_{\tau }}\r^{(0)}~,
\la{alp3}
\eeq
where $\l \al^{-1}_s\r^0=\fr{1}{7}\l 12\al^{-1}_2-5\al^{-1}_1+ 
12\De_2-5\De_1-7\De_3 \r $ corresponds to the value of $\al_s$ obtained
for MSSM (or MSSU5). The prime on $\al_s$ indicate that it is calculated
ignoring two loop effects coming from $\de_a$ terms.

Employing the reasonable asymptotic relations

\beq
\fr{m_b^{(0)}}{m_{\tau }^{(0)}}=1~,~~~
\fr{m_s^{(0)}}{m_{\mu }^{(0)}}=\fr{1}{3}~,~~~
\fr{m_d^{(0)}}{m_e^{(0)}}=3~,
\la{asym}
\eeq
and using $\l \al^{-1}_s\r^0=1/0.126$ \cite{alpsu5}, from
(\ref{alp3}) we get
$\l \al_s\r'\simeq 0.12$. 
Taking account of $\de_a$ terms, we have 

\beq
\al_s^{-1}=\l \al_s^{-1}\r'+\de~, 
\la{als2loop}
\eeq
where $\de=\fr{1}{7}(12\de_2-5\de_1-7\de_3)$.
In order to calculate $\de_a$ in (\ref{twoloop}), we have to know the masses
of doublet and triplet vectorlike states. From (\ref{matd}),
(\ref{matl}) and (\ref{relbigdets}), it is natural to assume that for each
family we
have $M_i^{d^c}\simeq M_G\lam_d^i$. Also for each family we will assume relation
(\ref{relmass}) which, taking into account (\ref{asym}), gives
$M_3^l=2M_3^{d^c}$, $M_2^l=2M_2^{d^c}/3$, $M_1^l=6M_1^{d^c}$.
Recall that for the PGB $SU(6)$ scenario, the preferred value of $\tan \bt$
is order unity \cite{pgbsu62}-\cite{pgbsu3}, so that 
$\lam_b\sim \lam_{\tau }\sim 10^{-2}$.
We also have the measured hierarchies between down quark Yukawa couplings: namely,
$\lam_d :\lam_s: \lam_b \sim \ep^5:\ep^2:1$, where $\ep\simeq 0.2$. Taking all this
into account,
for the mass spectra of the vectorlike states, it is quite natural to have:

$$
M_3^{d^c}=10^{-2}M_G~,~~~M_2^{d^c}=10^{-2}M_G\ep^2~,~~~
M_1^{d^c}=10^{-2}M_G\ep^5~,
$$
\beq
M_3^l=2\cdot 10^{-2}M_G~,~~~M_2^l=\fr{2}{3}10^{-2}M_G\ep^2~,~~~
M_1^l=6\cdot 10^{-2}M_G\ep^5~.
\la{spectra}
\eeq
In (\ref{twoloop}) we have

\beq
b_a^{\rho }=b_a+mb_a^{d^c}+nb_a^l~,~~~~
b_{ab}^{\rho }=b_{ab}+mb_{ab}^{d^c}+nb_{ab}^l~,
\la{brho}
\eeq
where $m$ and $n$ denote how many vectorlike triplet and doublet states
respectively we have at the appropriate mass scale. $b_a$,
$b_a^{d^c}$ and $b_a^l$ are given in (\ref{bfacs}), while

\begin{equation}
\begin{array}{cc}
b_{ab} \hs{-1mm}=\!\!\!\!\hs{-1mm} & 
\left (\begin{array}{ccc}
\hs{-1mm} \fr{199}{25}~,
&\fr{27}{5}~, &\fr{88}{5}
\\
\hs{-1mm} \fr{9}{5}~,
&25~,&24
\\
\hs{-1mm}\fr{11}{5}~,
&9~,&14
\end{array}\hs{-1.5mm}\right)\hs{0.1mm},
\end{array}
\begin{array}{cc}
b_{ab}^{d^c}\hs{-1mm}=\!\!\!\!\!\hs{-1mm} &{\left(\begin{array}{ccc}
\hs{-1mm} \fr{8}{75}~, &0~,&\fr{32}{15}
\\
\hs{-1mm} 0~, &0~,&0
\\
\hs{-1mm} \fr{4}{15}~, &0~,&\fr{34}{3}

\end{array}\hs{-1.5mm}\right)\hs{0.1mm},
}
\end{array}
\begin{array}{cc}
b_{ab}^l\hs{-1mm}=\!\!\!\!\!\hs{-1mm} &{\left(\begin{array}{ccc}
\hs{-1mm} \fr{9}{25}~,  &\fr{9}{5}~,& 0
\\
\hs{-1mm} \fr{3}{5}~, &7~, &0
\\
\hs{-1mm} 0~, &0~, & 0

\end{array}\hs{-1.5mm}\right)\hs{0.1mm}.
}
\end{array}
\label{bb}
\end{equation}
{}From (\ref{twoloop}), taking into account (\ref{spectra})-(\ref{bb}), 
we obtain $\de=0.056$, and according to (\ref{als2loop})
$\al_s(M_Z)=0.119$
in excellent agreement with the experimental data \cite{data}.
Numerical calculations confirm these estimations. The unification 
picture of gauge couplings is presented on Fig. 1.

As far as the up-type quark sector is concerned, the relevant couplings for
their mass generation are 
$\fr{\ga_{\al \bt }}{M^2} 15_{\al }15_{\bt }\Si H^2$, where $\al, \bt $
are
family indices and $M$ is some cutoff mass scale of the order of $M_G$. These
operators could emerge through exchange of some additional states
with masses $\sim M$ \cite{pgbsu62}.

As we have demonstrated, a realistic fermion mass
pattern is realized when $\lan \Si \ran $ aligns along the 
$SU(4)\times SU(2)\times U(1)$ direction (\ref{421dir}). This
indeed also happens to be the VEV direction required for
realization of the PGB mechanism
within $SU(6)$.We
refer the reader to \cite{pgbsu61, pgbsu62}, where detailed studies of
this question are presented.

In conclusion, let us note that in the present scenario,
it is possible to invoke flavor symmetries, the
simplest being ${\cal U}(1)$ for a natural
understanding of hierarchies between charged fermion masses and the CKM matrix
elements. The various neutrino oscillation scenarios are considered in
\cite{anu1}. If the flavor ${\cal U}(1)$ turns out to be anomalous, 
it also helps in achieving an `all order' DT hierarchy (see last two
refs. in \cite{pgbsu62}).  

\section*{Acknowledgements}

\noindent  
The work of Q. Shafi was supported in part by The US Department of 
Energy Grant No. DE-FG02-91ER40626.  

\bibliographystyle{unsrt}

\newpage

\vspace{-1cm}
\begin{figure}[tb]
\epsfysize=5.1in
\epsffile[105 220 500 500]{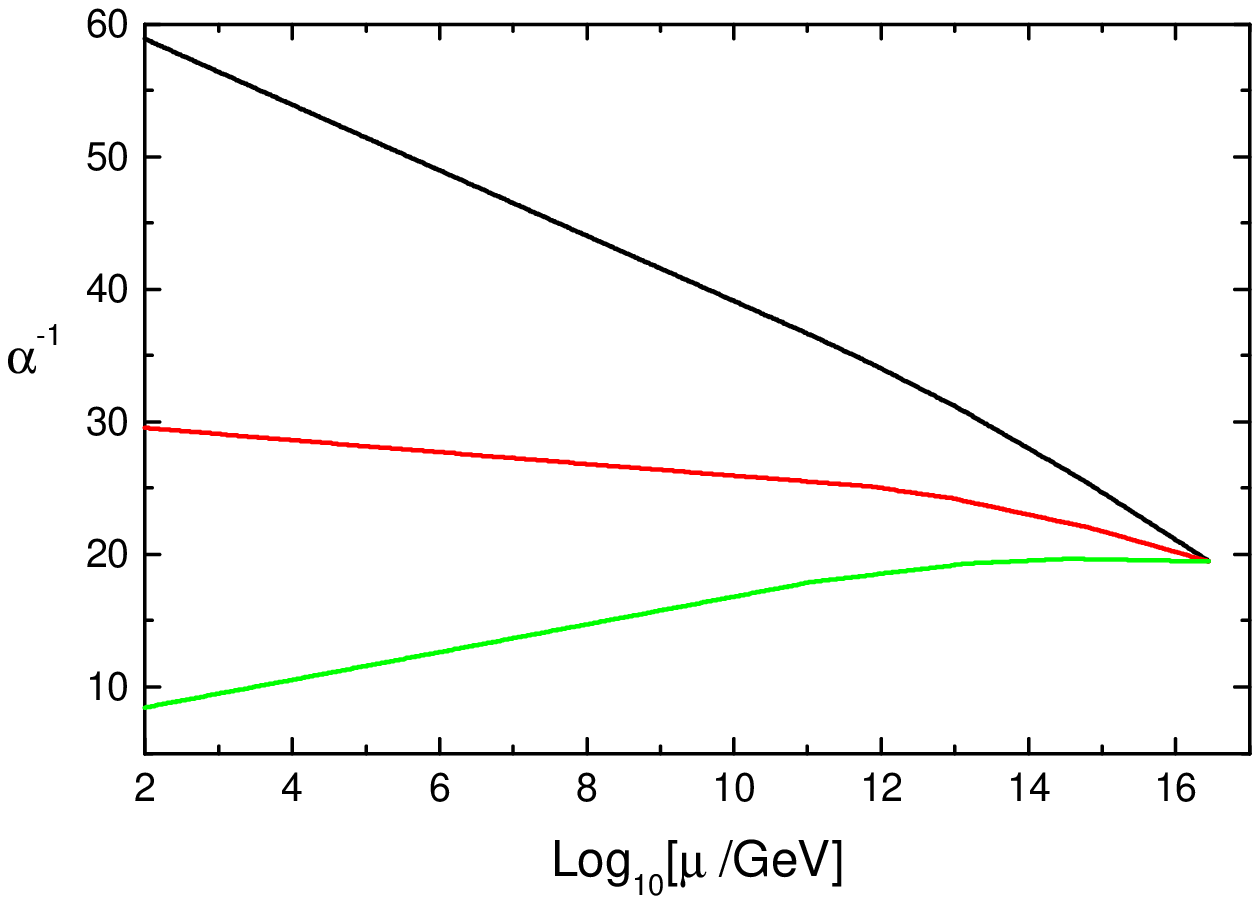} 
\begin{center}
\vspace{-1cm} 
\caption[]{$SU(6)$ Unification with $\al_s(M_Z)=0.119$,
$M_G=2.7\cdot 10^{16}$~GeV and $\al_G\simeq 1/19.5$.
}
\end{center}
\end{figure}

\end{document}